\documentclass{article}
\usepackage[pdftex]{color,graphicx}
\usepackage{amsmath,enumerate, amsfonts, ulem, url}
\usepackage{natbib}
\newcommand{\bm}[1]{\mbox{\boldmath{$#1 $}}}

\begin{document}
\baselineskip 18pt
\begin{center}
{\LARGE\textbf{Varying-coefficient functional additive models}}
\end{center}
\begin{center}
{\large Hidetoshi Matsui}
\end{center}

\begin{center}
\begin{minipage}{14cm}
{
\begin{center}
{\it {\footnotesize 
Faculty of Data Science, Shiga University \\
1-1-1, Banba, Hikone, Shiga, 522-8522, Japan. \\
}}

\vspace{2mm}

{\small hmatsui@biwako.shiga-u.ac.jp}
\end{center}
\vspace{1mm} 

{\small {\bf Abstract:} 
	We extend the varying coefficient functional linear model to the nonlinear model and propose a varying coefficient functional additive model.  
	The proposed method can represent the relationship between functional predictors and a scalar response where the response depends on an exogenous variable.
	It captures the nonlinear structure between variables and also provides interpretable relationship of them.  
	The model is estimated through basis expansions and penalized likelihood method, and then the tuning parameters included at the estimation procedure are selected by a model selection criterion.  
	Simulation studies are provided to show the effectiveness of the proposed method.  
	We also apply it to the analysis of crop yield data and then investigate how and when the environmental factor relates to the amount of the crop yield.  
}

\vspace{3mm}

}
\end{minipage}
\end{center}

\section{Introduction}
Functional data analysis (FDA) is a widely applicable technique for analyzing longitudinally observed data, and is applied in various fields of data, such as bioinformatics, medicine and meteorology \citep{RaSi2005,KoRe2017}.  
In particular, functional regression analysis is one of the most useful techniques in functional data analysis.  
The basic idea behind functional regression analysis is to treat longitudinally observed data for predictors and/or responses as smooth functional data, and then to elucidate the relationship between them from estimated model and to predict the newly observed data.  

There are various kinds of functional regression models according to the data structure of the predictor and the response.  
The most basic model is a functional linear model for a functional predictor and a scalar response, which is discussed in \cite{RaDa1991}, \cite{CaFeSa1999}, and \cite{GoFeCr2010}.  
On the other hand, the functional linear model for a functional predictor and a functional response is also considered in \cite{RaDa1991}, \cite{YaMuWa2005b}, \cite{MaKaKo2009}.  
Numerous extensions and improvements of these models are reported; \cite{Mo2015} and \cite{ReGoSh_etal2017} extensively review several studies on functional regression models.   

In this work we consider the situation where the predictor is a function while the response is a scalar, but it depends on another variable.  
The motivating data come from crop yield data of multi-stage tomatoes.  
A plant of the multi-stage tomatoes grows for a long time in a year, and fruits are harvested daily.  
In general, the amount of the crop yield depends on environmental factors such as the temperature and the amount of solar radiation.  
In addition, these effects may differ for the season of the year for the multi-stage tomatoes.  
Therefore we treat the seasonal time as an exogenous variable.  
For the analysis of such type of data, the varying-coefficient functional linear model (VCFLM) by \cite{CaSa2008} and \cite{WuFaMu2010} can be applied by treating the seasonal time as an exogenous variable.  
The VCFLM is an extension of the varying-coefficient model \citep{HaTi1993,HoRiWu_etal1998} to the functional linear model framework, and it can represent the relationship between a functional predictor and a scalar response varying with the exogenous variable.  
Specifically, we can interpret the relationship by investigating coefficient functions of the model.  
Several refinements and application of the VCFLM are discussed in \cite{PeZhTa2016}, \cite{LiHuZh2017}, and \cite{DaMaBa2018}.  
However, the VCFLM captures only the ``linear" relationship at fixed exogenous variable.  
In the case of crop yield data, if the temperature is moderately high, the yield will be high, but if the temperature is too high, the yield will decrease.  
It is difficult for the VCFLM to capture such relationship.  

To solve this problem, we extend the VCFLM to the nonlinear model for a continuous response variable.  
On of the extensions of the traditional linear model to the nonlinear one is an additive model \citep{HaTi1990}, and there are several extensions of the functional linear model to additive model frameworks.  
\cite{MuYa2008} proposed a functional additive model (FAM) that extend the linear term to the nonlinear one.  
Their approach uses Karhunen-Lo\'eve (KL) expansions and local polynomial regression, whereas \cite{McHoSt_etal2014} directly applies the basis expansion technique to the nonlinear structure, and \cite{FaJaRa2015} assumes an unknown link function between the linear predictor and the response.  
Among them, it is easier to interpret the \cite{MuYa2008}'s model in the viewpoint that how the functional predictor affect the response.  
In addition, \cite{MuWuYa2013}, \cite{ZhYaZh2014}, \cite{HaMuPa2018}, and \cite{WoLiZh2019} developed the FAM to several situations.  
\cite{IvStSc_etal2015}, \cite{ScGr2016}, and \cite{ScGeGr2016} considered comprehensive functional regression model including functional additive models.  

Using these ideas, we propose a novel varying-coefficient functional additive model (VCFAM).  
The VCFAM captures nonlinear relationship between functional predictors and a scalar response, where the response depends on an exogenous variable.  
Furthermore, we can interpret the relationship from the estimated model.  
We consider estimating the VCFAM by the penalized likelihood method along with the basis expansions.  
In order to select tuning parameters included in the penalty, we apply a model selection criterion for evaluating the estimated model, using the idea of \cite{KoKi2008}.  
Simulation studies are conducted to evaluate the effectiveness of the proposed method.  
Then we apply the VCFAM to the analysis of crop yield data of multi-stage tomatoes to investigate how the temperature during the cultivation affect the crop yield.  
We also consider predicting future yields using the past temperature and yield data.  

This paper is organized as follows.  
Section 2 briefly reviews existing models that relate to the proposed method, and then we introduce a VCFAM.  
Section 3 shows the method for estimating and evaluating the VCFAM.  
Simulation studies are given in Section 4, and then real data analysis are discussed in Section 5.  
Conclusions about our work are summarized in Section 6.  

\section{Model}
Before introducing our model, we overview some existing models; the functional linear model, the functional additive model and the varying-coefficient functional linear model.  
Then we propose a novel varying-coefficient functional additive model. 
\subsection{Existing models}
Suppose we have $n$ sets of a functional predictor and a scalar response $\{x_i(s), y_i; i=1,\ldots, n, s\in\mathcal{S}\in\mathbb{R}\}$, where $x_{i}(s)$ and $y_i$ are a functional predictor and a scalar response respectively.  
We also assume that the functional predictor $x_{i}(s)$ is expressed by truncated Karhunen-Lo\'eve (KL) expansions \citep{YaMuWa2005a};
\begin{align*}
x_{i}(s) = \sum_{k=1}^{q} \xi_{ik}\phi_{k}(s),
\end{align*}
where $\xi_{ik}$ and $\phi_{k}(s)$ $(k=1,\ldots, q)$ are functional principal component (FPC) scores and corresponding eigenfunctions with eigenvalues $\lambda_{k}$, and $q$ is the truncation number of principal components. 
The FPC scores satisfies $E(\xi_{ik}) = 0$ and $E(\xi_{ik}^2) = \lambda_k$,  and the eigenvalues satisfies $\lambda_{1} > \cdots >  \lambda_{q} >0$.   
In addition, the eigenfunctions are orthonormal basis, that is, $\int \phi_{k}(s)\phi_l(s)ds = \delta_{kl}$, where $\delta_{kl}$ is a Kronecker's delta.   
Although we can apply the well-known basis functions such as splines or radial basis functions \citep{GrSi1994} for $\{\phi_k(s)\}$, the KL expansion can represent data with smaller number of basis functions.  

The traditional functional linear model (FLM) is given in the form of 
\begin{align}
	y_i = \beta_0 + \int_{\mathcal S} x_i(s)\beta_1(s)ds + \varepsilon_i,
	\label{eq:FLM}
\end{align}
where $\beta_0$ is an intercept, $\beta_1(s)$ is a coefficient function and $\varepsilon_i$ are errors independently and identically distributed with mean zero and unknown variance. 
If we assume that the functional predictor and coefficient function are expressed by basis expansions (including KL expansion), the problem of estimating the model becomes that of estimating the ordinal linear model with predictors $\xi_{ik}$ \citep{RaSi2005}.  

The functional additive model (FAM) by \cite{MuYa2008} is given by
\begin{align}
	y_i = \sum_{k=1}^{q} f_k(\xi_{ik}) + \varepsilon_i,
	\label{eq:FAM}
\end{align}
where $f_k$ are unknown functions.   
The mean structure of the traditional functional linear model is transformed into the ordinal linear model with predictors $\xi_{ik}$ \citep{YaMuWa2005b}, whereas the FAM (\ref{eq:FAM}) is a natural extension to the additive model. 
Therefore we can capture more complex relationship between the predictor and the response.   


If the response depends on the exogenous variable $t_i,$ the following varying-coefficient functional linear model (VCFLM) is considered \citep{CaSa2008,WuFaMu2010};
\begin{align}
y_i = \beta_0(t_i) + \int_{\mathcal T} \beta_1(s, t_i)x_i(s)ds + \varepsilon_i, 
\label{eq:VCFLM}
\end{align}
where $\beta_0(t)$ is a baseline function and $\beta_1(s, t)$ is a coefficient surface.  
Then we can represent the relationship between the response and the predictor with varying $t$.  

\subsection{Varying-coefficient functional additive model}
Again we denote $n$ sets of observations as $\{x_{i}(s), y_i, t_i; s\in\mathcal S\subset\mathbb R \}$, where the response $y_i$ depends on the exogenous variable $t_i$ as well as a functional predictor $x_i(s)$.  
In addition, $x_{i}(s)$, $y_i$ are supposed to be centered so that $\sum_{i=1}^n x_i(s) = 0$ and $\sum_{i=1}^n y_i = 0$.  
To express the relation of these variables, we extend the VCFLM (\ref{eq:VCFLM}) to the additive model framework (\ref{eq:FAM}).  
When applying the functional additive model, \cite{ZhYaZh2014} proposed transforming the functional principal component (FPC) scores $\lambda_{k}$ into $\zeta_{k} \in [0,1]$ by using some monotonic function such as the cumulative distribution function $\Phi(x; 0, \lambda_k)$ of the normal distribution $N(0, \lambda_{k})$. 
That is , $\zeta_{k}$ is given by $\zeta_{k} = \Phi(\xi_{k}; 0, \lambda_{k})$.

Using these ideas, we model the relationship between the response and predictors as the following varying-coefficient functional additive model (VCFAM);
\begin{align}
	y_i = \sum_{k=1}^{q} f_{k}(\zeta_{ik}, t_i) + \varepsilon_i.  
	\label{eq:VCFAM}
\end{align}
where $f_{k}$ is a nonlinear function of $\zeta_{ik}$ and $t_i$, here we assume that $f_{k}$ satisfies $E[f_k(\zeta_{ik}, t_i)|t_i] = 0$.  
In addition, $\varepsilon_i$ is an error that $\bm{\varepsilon} = (\varepsilon_1,\ldots, \varepsilon_n)^T$ follows normal distribution with mean vector $\bm 0$ and variance covariance matrix $\Sigma$.
In our application described in Section 5 the index $i$ corresponds to the observed time, so we assume that $\varepsilon_i (i=1\ldots, n)$ depends on each other rather than i.i.d.
Advantages of the VCFAM (\ref{eq:VCFAM}) is that we can consider the nonlinear relationship between the response and predictors at varying $t$.  

The VCFAM (\ref{eq:VCFAM}) can be extended to the situation where there are multiple functional predictors $\{x_{i1}(s),\ldots, x_{ip}(s)\}$;
\begin{align*}
	y_i = \sum_{j=1}^p \sum_{k=1}^{q_j} f_{jk}(\zeta_{ijk}, t_i) + \varepsilon_i,  
	\label{eq:mVCFAM}
\end{align*}
where $f_{jk}(\cdot, \cdot)$ are nonlinear functions and $\zeta_{ijk}$ are derived from FPC scores by the same strategies as $\zeta_{ik}$.  

\section{Estimation}
In order to estimate the unknown functions $f_{k}$ in the VCFAM (\ref{eq:VCFAM}), we assume that this is expressed by basis expansions as follows.
\begin{align*}
f_{k}(\zeta_{ik}, t_i) 
&= \sum_{h=1}^{m_{1}}\sum_{l=1}^{m_{2}} \eta_{kh}(\zeta_{ik}) \theta_{khl} \psi_{kl}(t_i) 
= \bm{\eta}_{k}(\zeta_{ik})^T\Theta_{k}\bm{\psi}_{k}(t_i),
\end{align*}
where $\bm\eta_{k}(\zeta) = (\eta_{k1}(\zeta),\ldots,\eta_{km_{1}}(\zeta))^T$ and $\bm\psi_{k}(t) = (\psi_{k1}(t),$ $\ldots,\psi_{km_{2}}(t))^T$ are vectors of basis functions and $\Theta_{k}=(\theta_{khl})_{hl}$ are $m_{1} \times m_{2}$ matrices of unknown parameters.  
Using this assumption, the VCFAM (\ref{eq:VCFAM}) can be expressed as
\begin{align*}
y_i 
&= \sum_{k=1}^{q} \bm{\eta}_{k}(\zeta_{ik})^T\Theta_{k}\bm{\psi}_{k}(t_i) + \varepsilon_i \\
&= \sum_{k=1}^{q} \left\{\bm\psi_{k}^T(t_i)\otimes \bm\eta_{k}^T(\zeta_{ik})\right\} {\rm vec}\Theta_{k} + \varepsilon_i,
\end{align*}
Then the VCFAM is given by 
\begin{align*}
\bm y 
&= \sum_{k=1}^{q}X_{k}{\rm vec}\Theta_{k} + \bm\varepsilon = X\bm\theta + \bm{\varepsilon},
\end{align*}
where 
\begin{gather*}
\bm y = \begin{pmatrix}
y_1 \\
\vdots \\
y_n
\end{pmatrix},~
X_{k} = \begin{pmatrix}
\bm\psi_{k}^T(t_1)\otimes \bm\eta_{k}^T(\zeta_{1k}) \\
\vdots \\
\bm\psi_{k}^T(t_n)\otimes \bm\eta_{k}^T(\zeta_{nk})
\end{pmatrix},~
\bm\varepsilon = \begin{pmatrix}
\varepsilon_1 \\
\vdots \\
\varepsilon_n
\end{pmatrix}, \\
X = \left(X_{1},\ldots, X_{q}\right), 
\bm{\theta} = \left(({\rm vec}\Theta_{1})^T, \ldots, ({\rm vec}\Theta_{q})^T\right),
\end{gather*}
and therefore the VCFAM (\ref{eq:VCFAM}) has a probability density function 
\begin{align*}
p(\bm y; \bm{\theta}, \Sigma) = \frac{1}{(2\pi)^{n/2}|\Sigma|^{1/2}}\exp\left\{
-\frac{1}{2}(\bm y - X\bm{\theta})^T\Sigma^{-1}(\bm y - X\bm{\theta})
\right\}. 
\end{align*}

The unknown parameter $\bm{\theta}$ is estimated by maximizing the penalized log-likelihood function given by 
\begin{align}
\ell_\lambda(\bm{\theta}) &= \log p(\bm y; \bm{\theta}, \Sigma) - n\bm{\theta}^T\Omega_\lambda\bm{\theta},
\label{eq:ploglike} 
\\
\Omega_\lambda &= I_q \otimes \left\{\lambda_{\zeta} (I_{m_2}\otimes P_{m_1}) + \lambda_t (P_{m_2}\otimes I_{m_1})\right\}, \nonumber
\end{align}
where $\lambda_{\zeta}, \lambda_{t}>0$ are regularization parameters and $P_{m_1}$ and $P_{m_2}$ are $m_1\times m_1$ $m_2\times m_2$ non-negative definite matrices, respectively.  
The matrix $\Omega_\lambda$ imposes penalties for the smoothness of the $f_k(\zeta, t)$ with respect to $\zeta$ and $t$ directions, and the amounts of penalties are controlled by $\lambda_\zeta$ and $\lambda_t$, respectively.  
By maximizing the penalized log-likelihood function (\ref{eq:ploglike}), a maximum penalized likelihood estimator of $\bm{\theta}$ is given by 
\begin{align*}
\hat{\bm\theta} = \left(X^T\Sigma^{-1}X + n\Omega_\lambda\right)^{-1}X^T\Sigma^{-1}\bm y,  
\end{align*}
where $\hat{\Sigma}$ is an estimator of $\Sigma$ and how to estimate it depends on the structure of $\Sigma$.  
For details, see, e.g. \cite{FaKnLa_etal2013}. 
Then we have a statistical model for the VCFAM by plugging the estimators $\hat{\bm{\theta}}$ and $\hat{\Sigma}$ into (\ref{eq:ploglike}).   

The VCFAM (\ref{eq:VCFAM}) estimated by the above method depends on tuning parameters such as the numbers $m_{1}$, $m_{2}$ of basis functions for $f_{jk}$ and the regularization parameters $\lambda_\zeta$, $\lambda_t$.  
In order to select appropriate values of them, we use an AIC-type model selection criterion \citep{Ak1974}.  
Using the result of \cite{HaTi1990}, the AIC for evaluating the statistical model is given by
\begin{align}
AIC = -2\log p(\bm y; \hat{\bm{\theta}}, \hat{\Sigma}) + 2\widehat{df},
\label{eq:IC}
\end{align}
where $\widehat{df}$ is an effective number of parameters obtained by $\widehat{df} = X(X^T\Sigma^{-1}X+n\Omega_\lambda)^{-1}X^T\Sigma^{-1}$.  
We select the values of the tuning parameters that minimize the AIC and treat the corresponding model as an optimal one.  
\section{Simulation}
We conduct simulation studies to investigate the effectiveness of the proposed method.  
Here we referred the setting of the simulation study to \cite{ZhYaZh2014}.  
First, we set the eigenvalues of the functional predictors by $\lambda_k = 45.25\times 0.64^k, k=1,...,q$, where the number of principal components is $q=20$.  
Next we computationally generated the $k$-th PC scores of the $i$-th subject $\xi_{ik}$ from $N(0, \lambda_k)$.  
Here we used Fourier series for the eigenfunctions $\{\phi_k(s)\}$.  
Then the longitudinal data for the predictor are generated by
\begin{align*}
x_{i\tau} = \mu(s_\tau) + \sum_{k=1}^{q}\xi_{ik}\phi_{k}(s_\tau) + e_{i\tau}, 
\end{align*}
where $\mu(s)$ is a mean function and here this is $\mu(s)=s+\sin(s)$, $s\in [0,1]$ and $e_{i\tau}\sim^{i.i.d} N(0, 0.2)$.  
Furthermore, $\tau = 1,\ldots, r$ with $r=21$ are the numbers of time points and we assume they are equally spaced on $[0,1]$ and are the same for individual.   

\begin{figure}[t]
	\begin{center}
		\includegraphics[width=0.45\hsize]{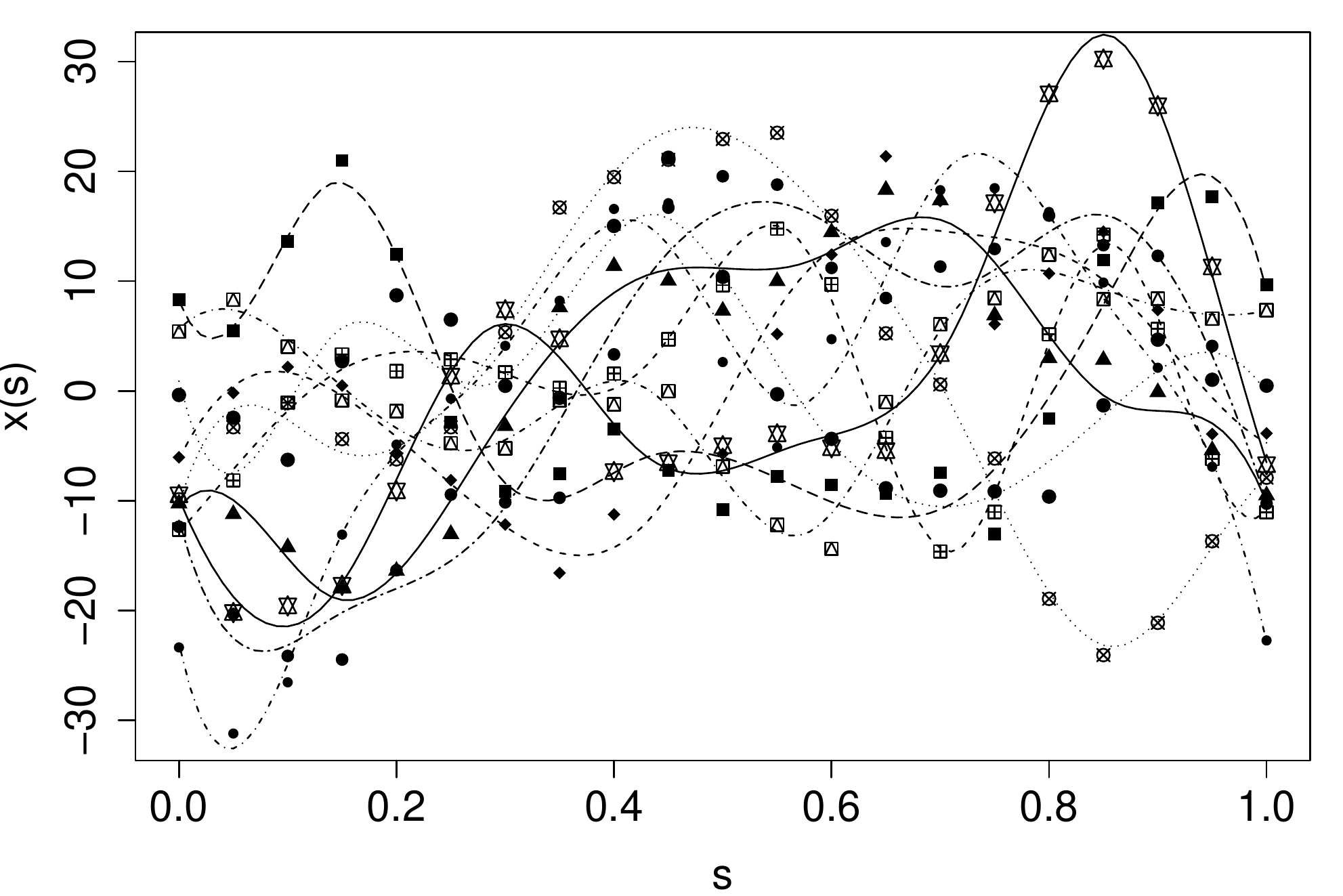}		
		\includegraphics[width=0.45\hsize]{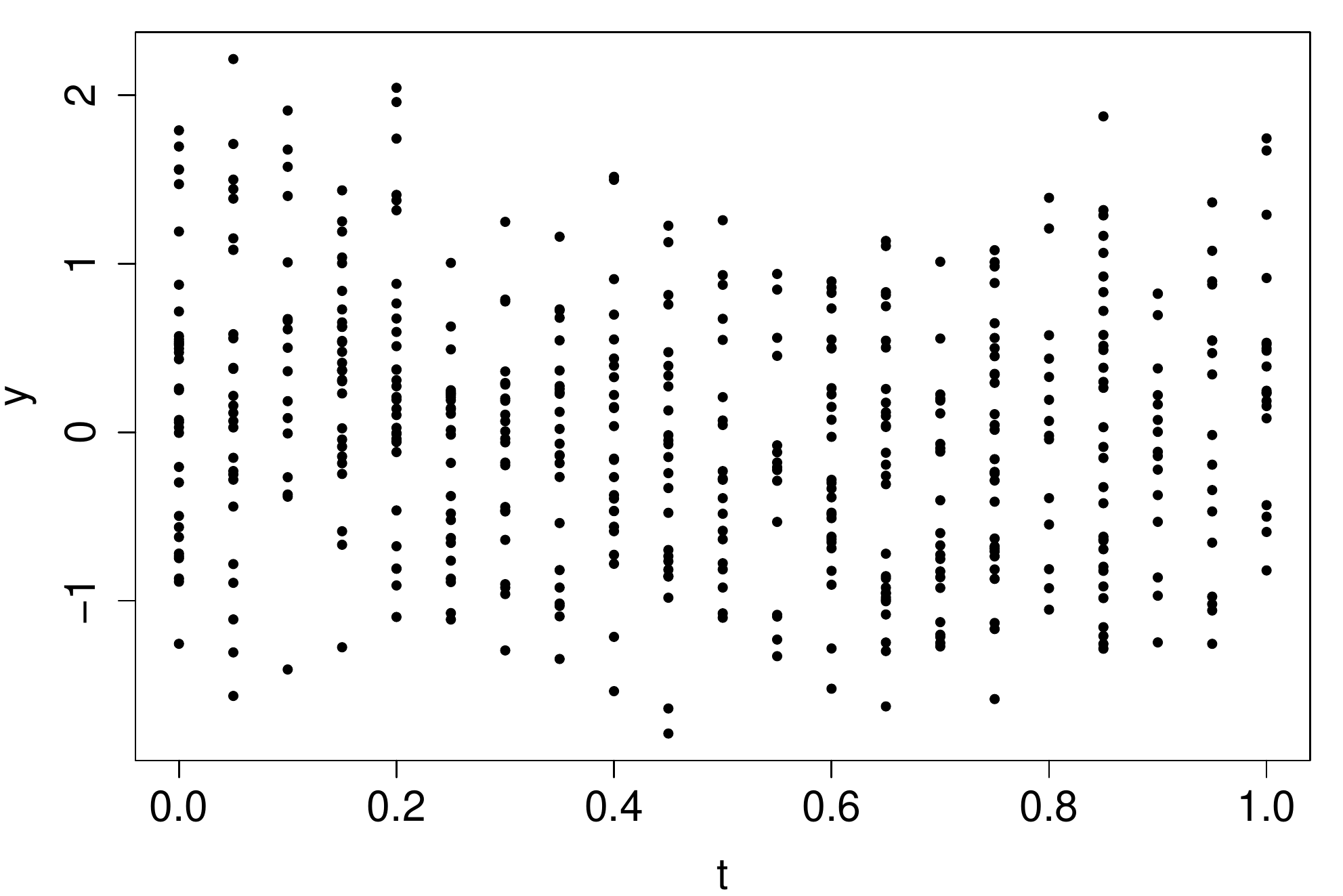}		
	\end{center}
	\caption{Example of the simulated data. Left: 10 examples for $x_{i\tau}$ and their functions. Right: 500 responses $y_i$ with varying $t$.  } 
	\label{fig:sim_data}
\end{figure}

We set true unknown functions $f_k(\zeta, t)$ as follows: 
\begin{align*}
f_1(\zeta, t) &= \cos \{\pi(\zeta + t)\},\\
f_2(\zeta, t) &= \sin\{2\pi(\zeta + t-1/2)\},\\
f_3(\zeta, t) &= \zeta^2 - 1/3, 
\end{align*}
and then the scalar response $y_i$ is given by
\begin{align*}
y_{i} = g_i + \varepsilon_i,~~
g_i = \sum_{k=1}^{q} f_k(\zeta_{ik}, t_i)
\end{align*}
where $\zeta_{ik} = \Phi(\xi_{ik}; 0, \lambda_k)$, $\varepsilon_i\sim^{i.i.d} N(0, (\sigma R_y)^2)$ with $R_y = \max_i(g_i) - \min_i(g_i)$ and a standard deviation parameter $\sigma$.   
By the above setting, we can obtain a simulated dataset $\{x_{i\tau}, y_i; i=1,\ldots, n, \tau=1,\ldots,r\}.$
Figure \ref{fig:sim_data} shows simulated data for the predictor and the response.  

For this dataset, we applied the proposed method and then estimated the unknown functions $f_k(\zeta, t)$ and evaluated the prediction accuracy.  
To do it, we transformed the data for predictor $\{x_{i\tau}\}$ into functional data $x_i(s)$, and then estimated the FPC scores $\xi_{ik}$ by applying the functional principal component analysis.  
Here we used R packages {\tt fda} for this process.  
We estimated parameter $\bm{\theta}$ of the VCFAM using the method described in Section 3.  
We fixed the numbers $m_1, m_2$ of basis functions $\bm\eta_{k}(\zeta)$ and $\bm\psi_{k}(t)$ to be 10 and 8 respectively for the computational simplicity, while the value of regularization parameter $\lambda_\zeta$ and $\lambda_t$ are selected by model selection criterion AIC (\ref{eq:IC}).  

We repeated this strategy for 100 times, and then calculated averages of 100 mean squared errors $MSE = \sum_{i=1}^n (g_i - \hat y_i)^2 / n$, where $\hat y_i = \sum_{k=1}^q \hat f(\zeta_{ik}, t_i)$.  
We compared the prediction accuracy of the proposed VCFAM with several other models; the VCFLM (\ref{eq:VCFLM}), FAM (\ref{eq:FAM}) with an additive nonlinear term for $t$ (denoted by FAM1), (\ref{eq:FAM}) itself (denoted by FAM2) and the functional linear model (FLM).

\begin{figure}[ht]
	\begin{center}
		\includegraphics[width=1\hsize]{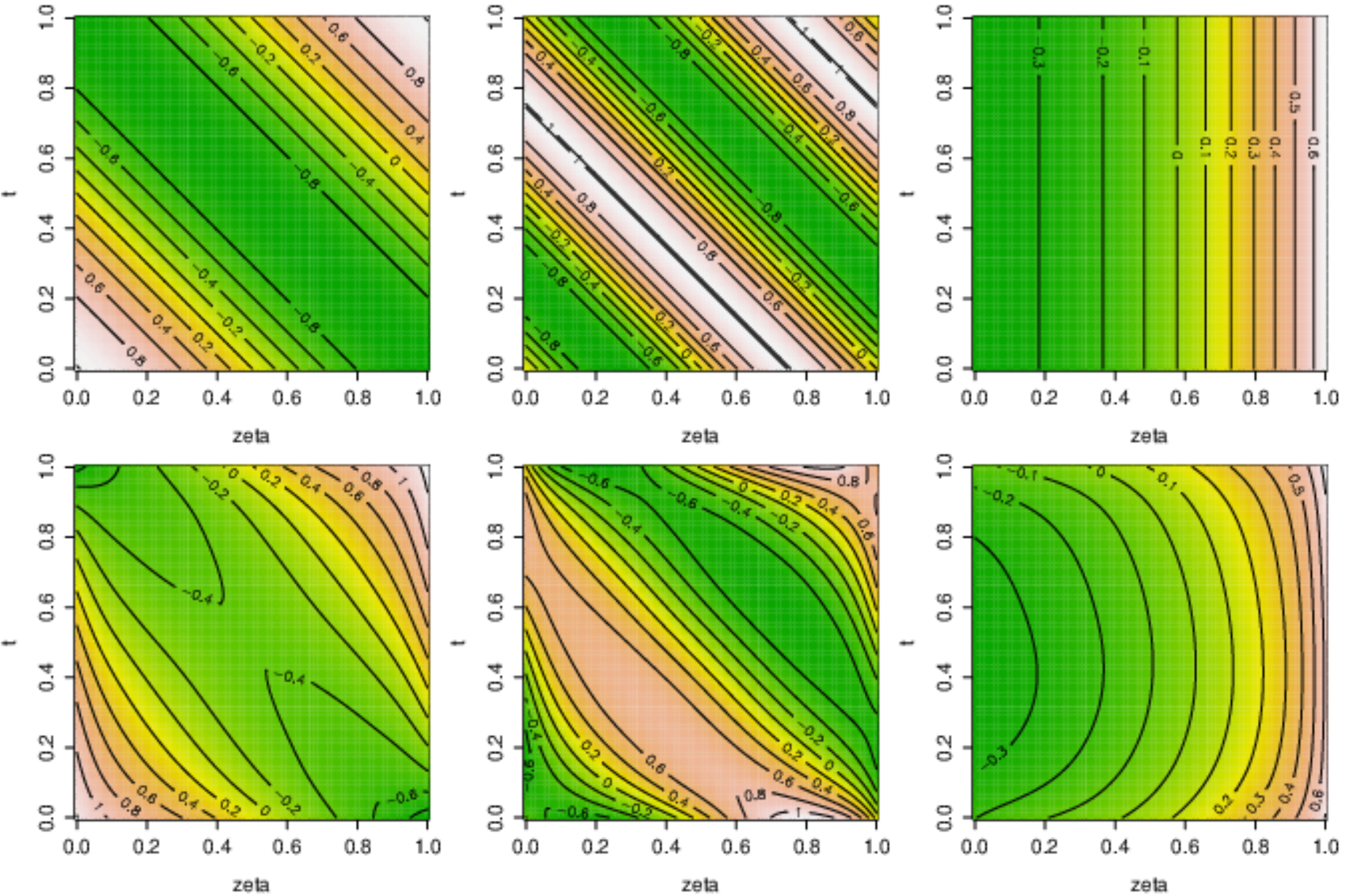}		
	\end{center}
	\caption{Top: True functions $f_k(\zeta, t)$ ($k=1,2,3$ from left to right). Bottom: Estimated functions $\hat f_k(\zeta, t)$ $(k=1,2,3)$.} 
	\label{fig:sim_fhat}
\end{figure}

Table \ref{tab:sim} shows results of the prediction.  
This shows that the proposed VCFAM minimizes the MSE for all cases, followed by VCFLM.  
FAM1 gives larger MSES compared to VCFAM and VCFLM, which indicates that the varying-coefficient model is more effective.  
Figure \ref{fig:sim_fhat} shows true nonlinear functions $f_k(\zeta, t)$ and estimated functions obtained by averaging 100 estimated surfaces $\hat f_k(\zeta, t).$ 
These figures indicate that our method roughly reconstructs the true functions.  

\begin{table}[t]
	\centering
	\caption{Results for simulation studies. Values in the tables show averaged MSEs $(\times 10)$ and their standard deviations $(\times 10^2)$ are given in parentheses.}
	\label{tab:sim}
	\begin{tabular}{rccccc}
		\hline
		& VCFAM & VCFLM & FAM1 & FAM2 & FLM \\ 
		\hline
		$n=500$  &&&&&\\
		$\sigma=0.05$ & 0.230 & 3.005 & 4.402 & 5.435 & 6.138 \\ 
				 & (1.247) & (1.880) & (2.734) & (3.194) & (3.326) \\ 
		$\sigma=0.1$  & 0.343 & 3.034 & 4.443 & 5.470 & 6.150 \\ 
			     & (1.401) & (1.682) & (2.664) & (3.650) & (3.859) \\ 
		\hline
		$n=1000$  &&&&&\\
		$\sigma=0.05$ & 0.147 & 3.063 & 4.438 & 5.413 & 6.124 \\ 
				 & (0.649) & (1.249) & (2.034) & (2.620) & (2.525) \\ 
		$\sigma=0.1$  & 0.207 & 3.098 & 4.450 & 5.421 & 6.108 \\ 
				 & (0.787) & (1.364) & (2.080) & (2.642) & (2.536) \\ 
		\hline
	\end{tabular}
\end{table}%
%
\begin{figure}[t]
	\begin{center}
		\includegraphics[width=0.45\hsize]{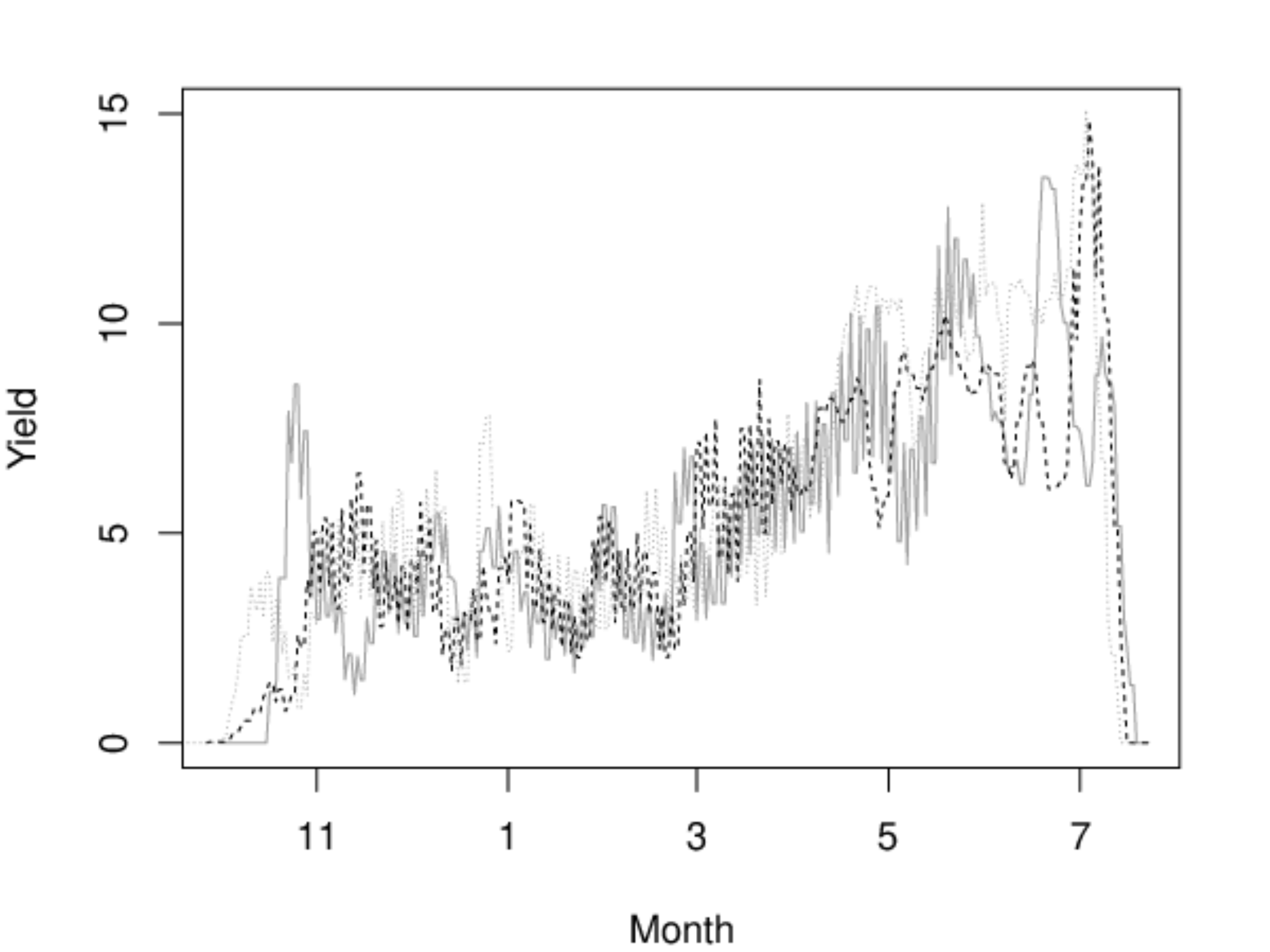}
		\includegraphics[width=0.45\hsize]{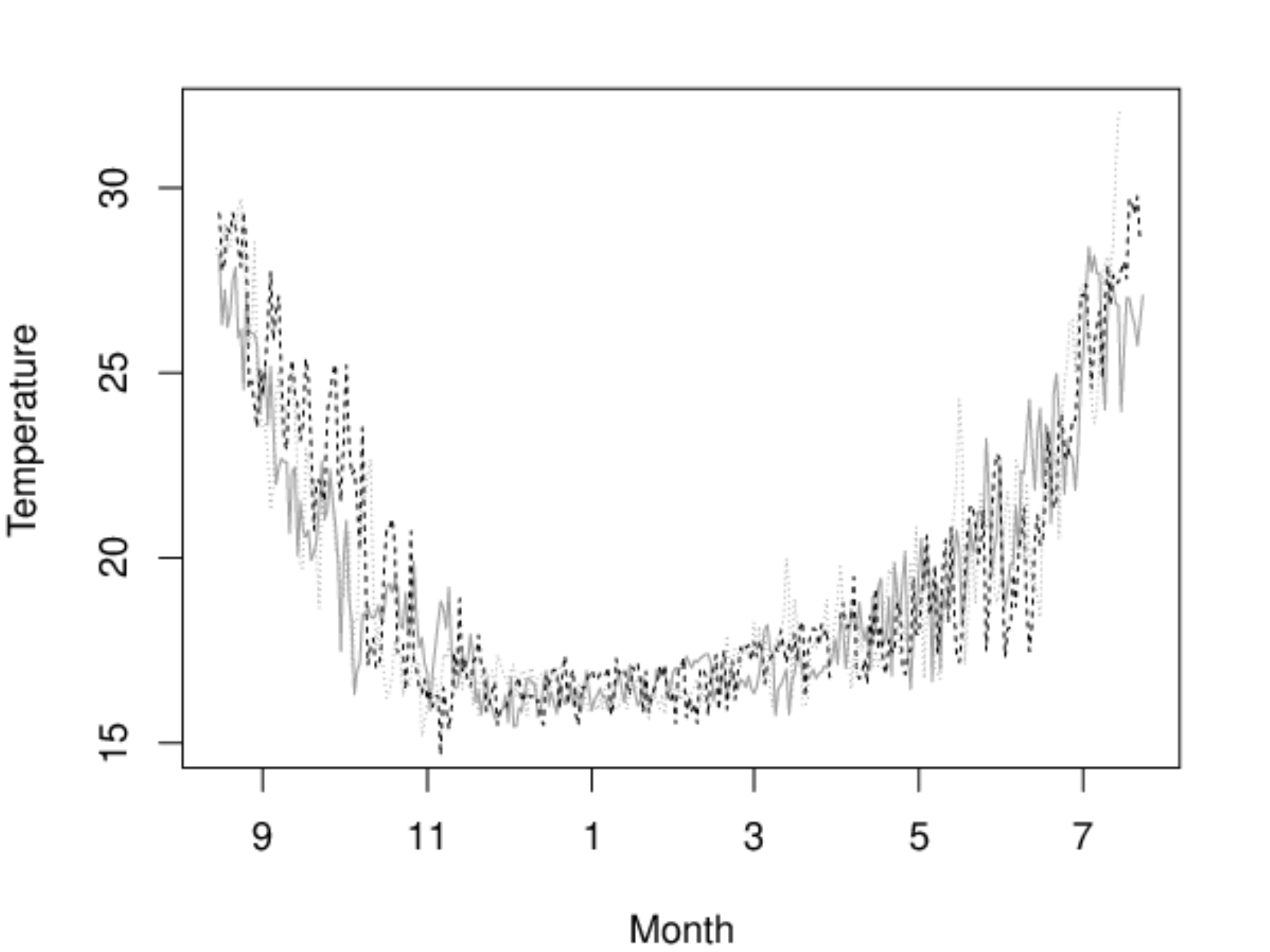}		
	\end{center}
	\caption{Left: amount of crop yield for 3 terms, where the moving averages for 7 days are shown. Right: daily temperatures for 3 terms during the cultivation term. } 
	\label{fig:tomato_data}
\end{figure}
\begin{figure}[ht]
	\begin{center}
		\includegraphics[width=0.6\hsize]{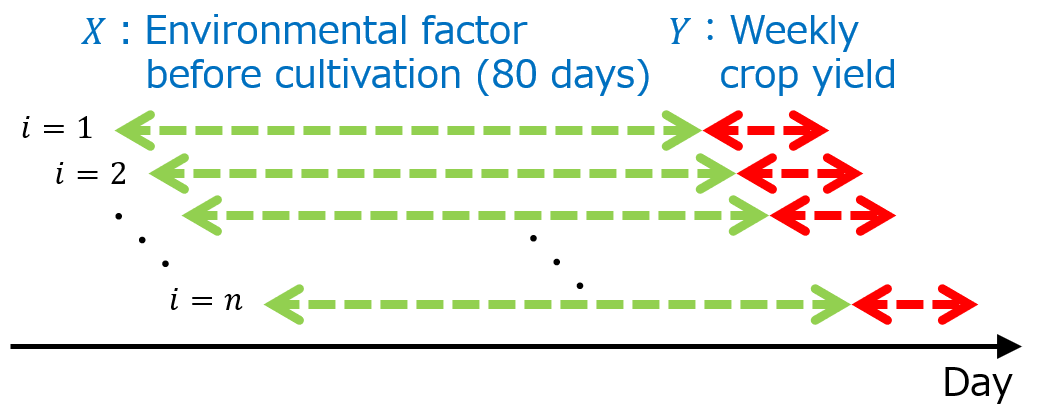}
	\end{center}
	\caption{Relation of the data for the analysis.} 
	\label{fig:setting}
\end{figure}
\section{Real data analysis}
We applied the proposed method to the analysis of the crop yield data for multi-stage tomatoes cultivated in a greenhouse on a farm in Kobe, Japan.  
Each seedling of the multi-stage tomatoes grows for about one year from August to next July, and are harvested almost every day from October to next July. 
In this study, we used daily yield data of single breed measured from October 2015 to July 2018, which is shown in Figure \ref{fig:tomato_data} left.  
However, the original data of the yield vary greatly because the yield is 0 when the farm is on a holiday, which makes the analysis too difficult.  
Therefore, we treat the moving averages of the yield up to 7 days from the harvest date as data.  
Environmental factors such as temperature and CO2 concentration are repeatedly measured by measuring equipment installed inside and outside the greenhouse, here we used the temperature inside the greenhouse as the data for environmental factor (Figure \ref{fig:tomato_data} right).  
It is considered that the growth of the tomato fruits are influenced by environmental factors during 80-day period before maturing the tomato.  
We used hourly averaged data for environmental factors as functional data.  
Therefore, we constructed a regression model, treating the daily yield of tomatoes as a response and the temperature corresponding to 80 days before the maturing day as a functional predictor.   
A set of an yield of certain day and 80-day temperature before the day corresponds to an individual, as shown in Figure \ref{fig:setting}, and the sample size is $n=836$.  
Readers may think that the analysis of this dataset can be applied by the function-on-function regression model with sparsely observed data discussed in \cite{YaMuWa2005b}.
In our case, however, the number of time points for the response is one for individual, and is not included in the function-on-function regression model.  
If we have the crop yield data for decades of years, we may be able to apply the function-on-function regression models by treating the yearly crop yield data as individuals, but the observed period of the dataset is only three years.  
For such dataset, our VCFAM is applicable.  

We transformed the data for the temperature into functional data and then calculated the FPCs, and then applied the VCFAM (\ref{eq:VCFAM}), where $s$ and $t$ correspond to the day before cultivation and the day of the year of the cultivation, respectively.  
The model is estimated by the penalized likelihood method and the tuning parameters are selected by AIC.  

\begin{figure}[t]
	\begin{center}
		\includegraphics[width=0.45\hsize]{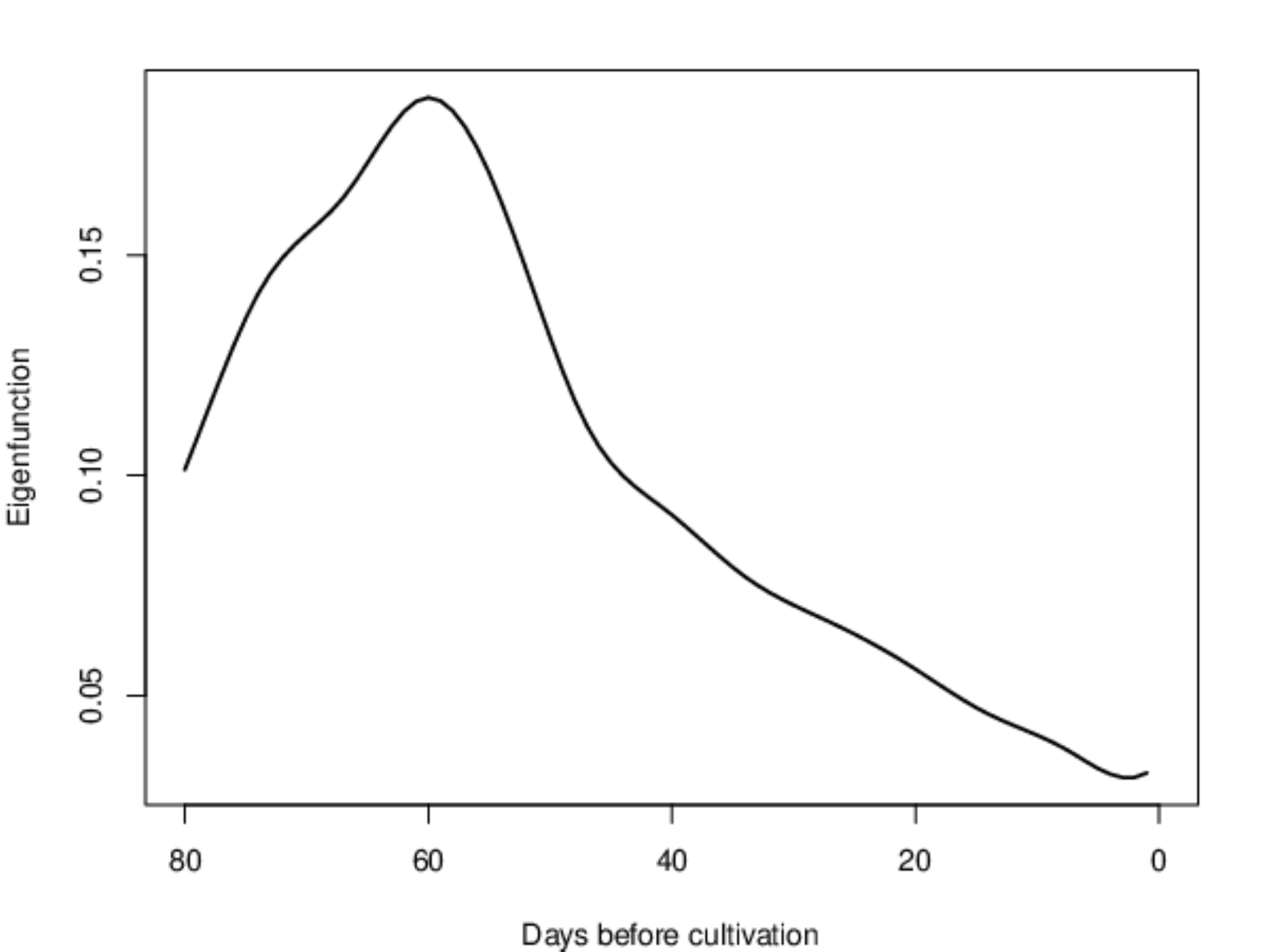}		
		\includegraphics[width=0.45\hsize]{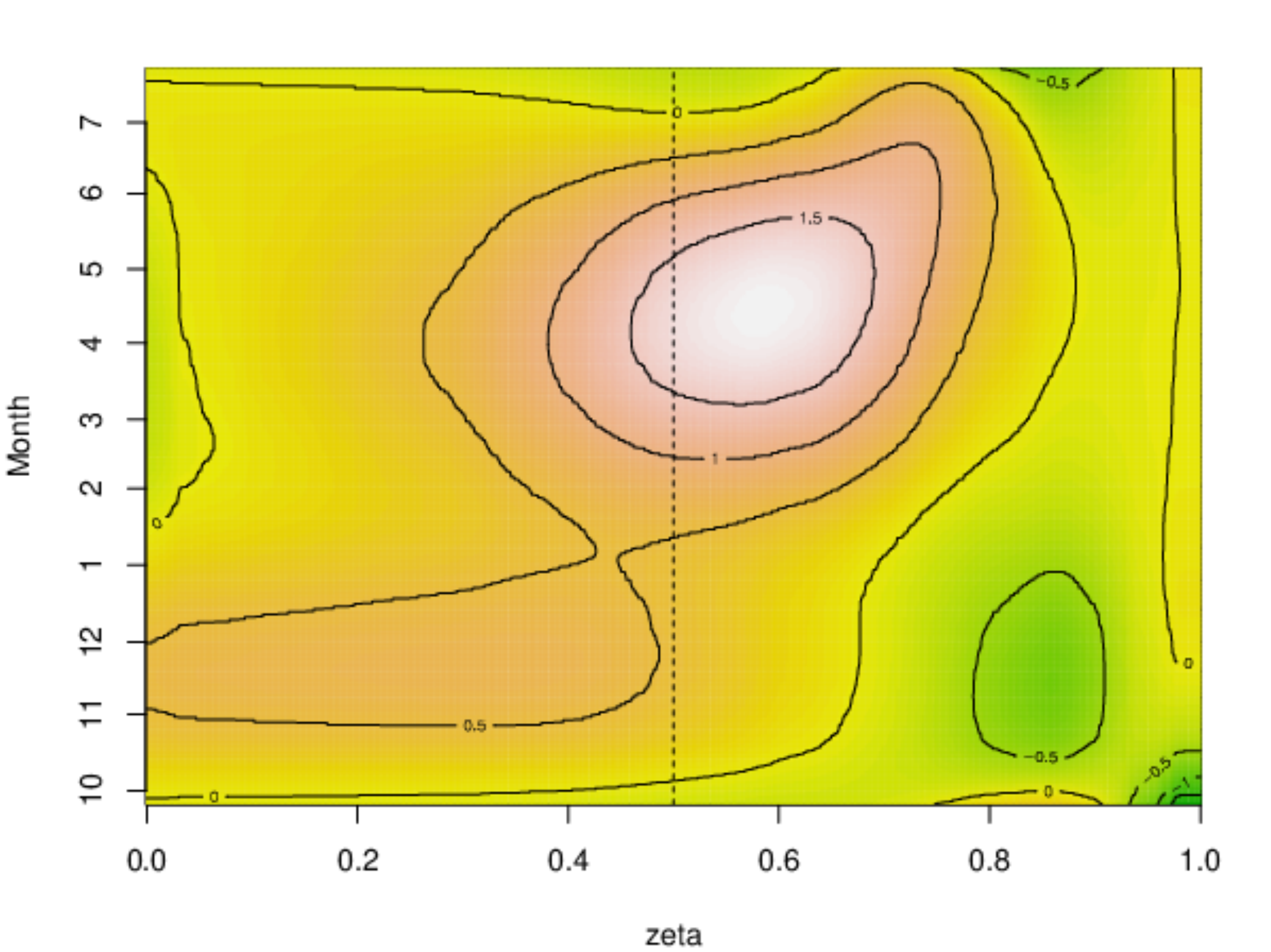} \\		
		\includegraphics[width=0.45\hsize]{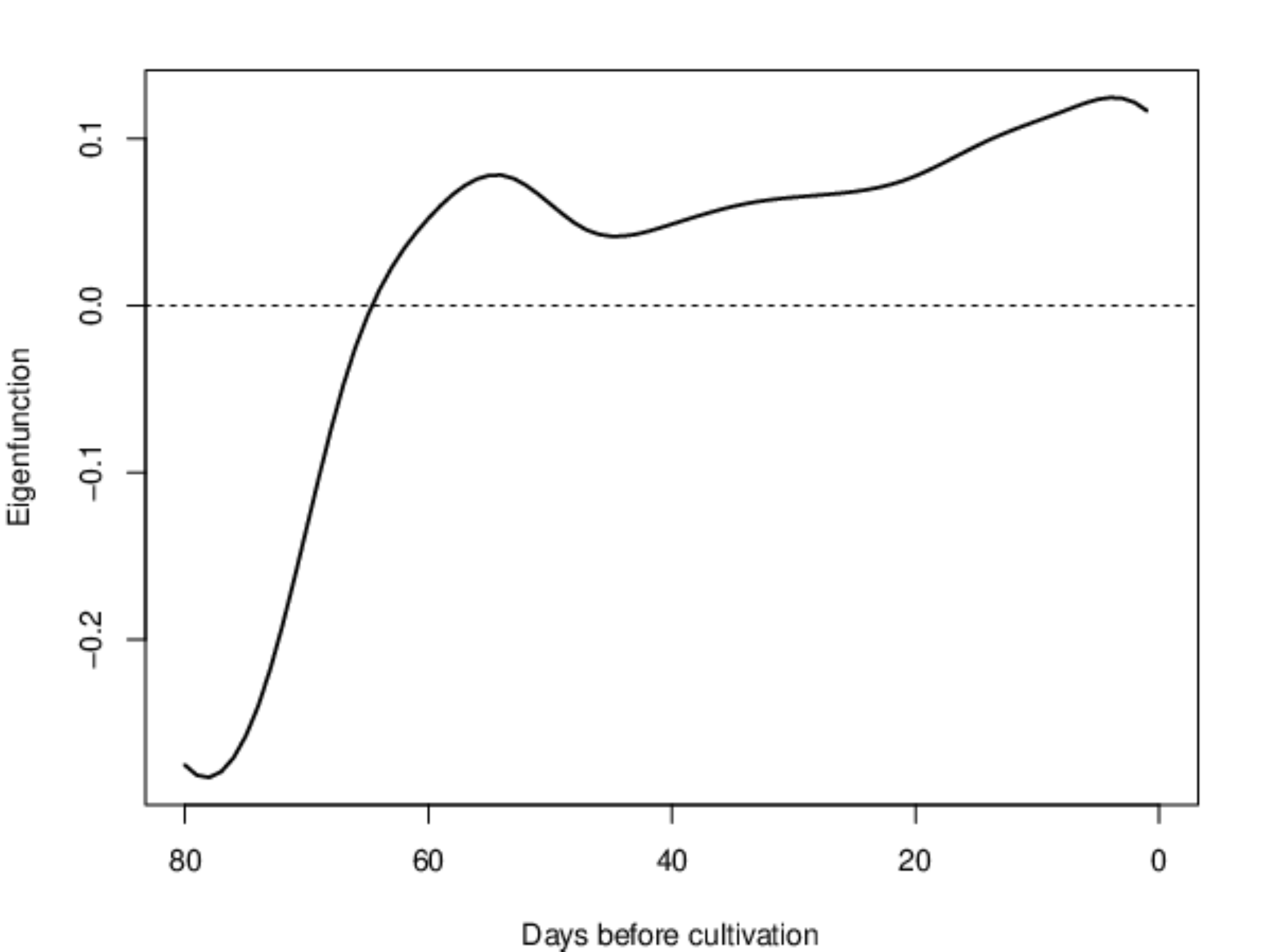}		
		\includegraphics[width=0.45\hsize]{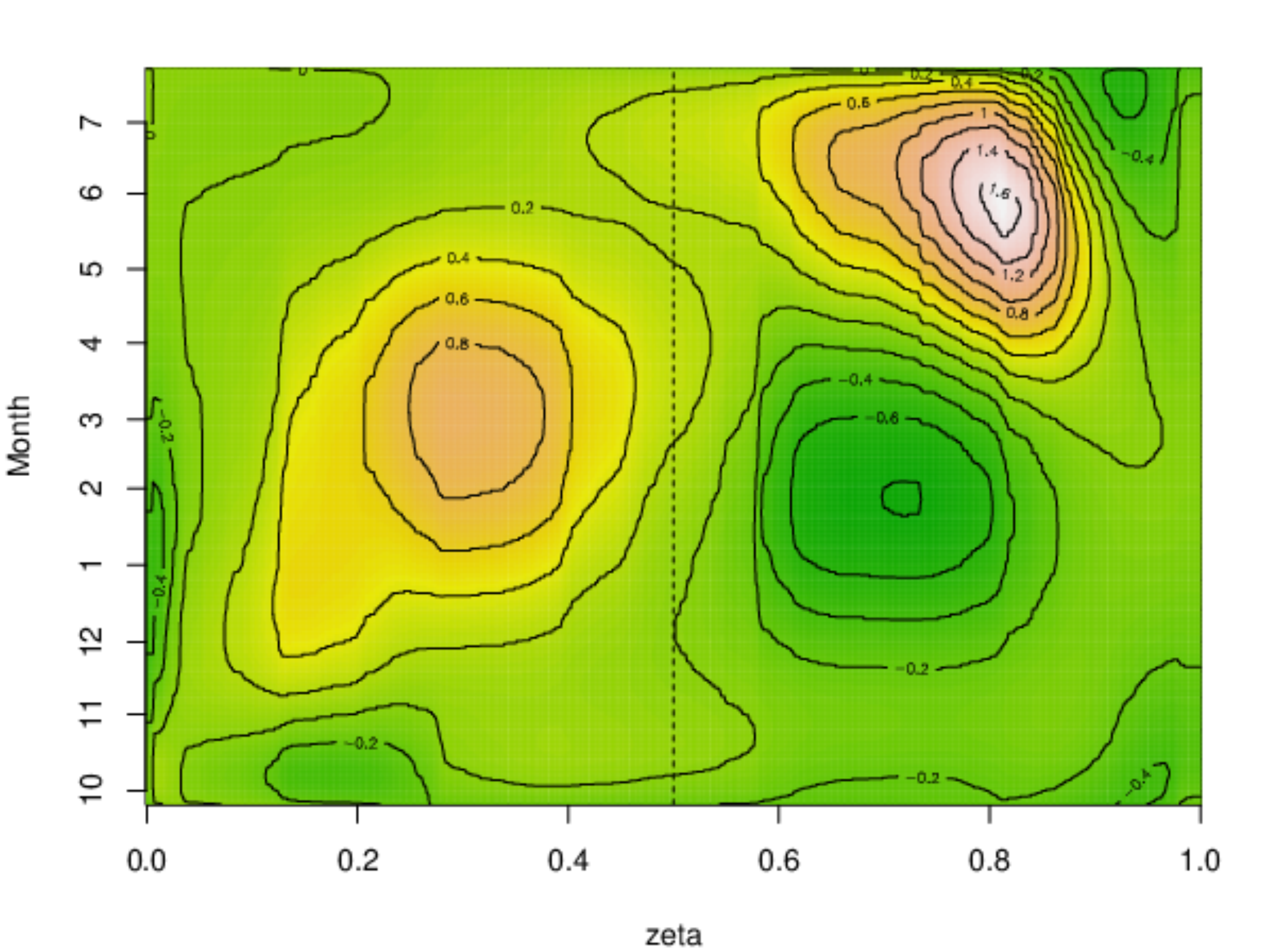}		
	\end{center}
	\caption{Left: estimated eigenfunctions $\phi_{jk}(s)$. Right: estimated functions $\hat f_k(\zeta, t)$.  Top figures correspond to the 1st FPC and the bottoms correspond to the 2nd FPC. } 
	\label{fig:tomato_vcfam}
\end{figure}

Figure \ref{fig:tomato_vcfam} shows estimated first and second eigenfunctions $\phi_{k}(s)$ $(k=1,2)$ of FPC and corresponding regression functions $f_{k}(\zeta, t)$.  
The eigenfunction for the first FPC means a high temperature especially at 60 days before cultivation, and corresponding $f_{k}(\zeta, t)$ shows when and how this effect is high.  
The surface in the top right of Figure \ref{fig:tomato_vcfam} shows positive at most area, which indicates that when the temperature at 60 days before cultivation is moderately, the crop yield is also high.  
In particular, in April, when temperature at that day before cultivation is moderately high, the crop yield increases. 
However, if the temperature is too high, the yield decreases.  
The second FPC shows the increase of the temperature in 80 days before cultivation.  
The corresponding regression function indicates that, for the crop yield in February and March, when the temperature decreases compared to 80 days before cultivation, the crop yield increases.  
On the other hand, in June, when the temperature increases compared to 80 days before cultivation, the crop yield increases.  

\begin{figure}[t]
	\begin{center}
		\includegraphics[width=0.8\hsize]{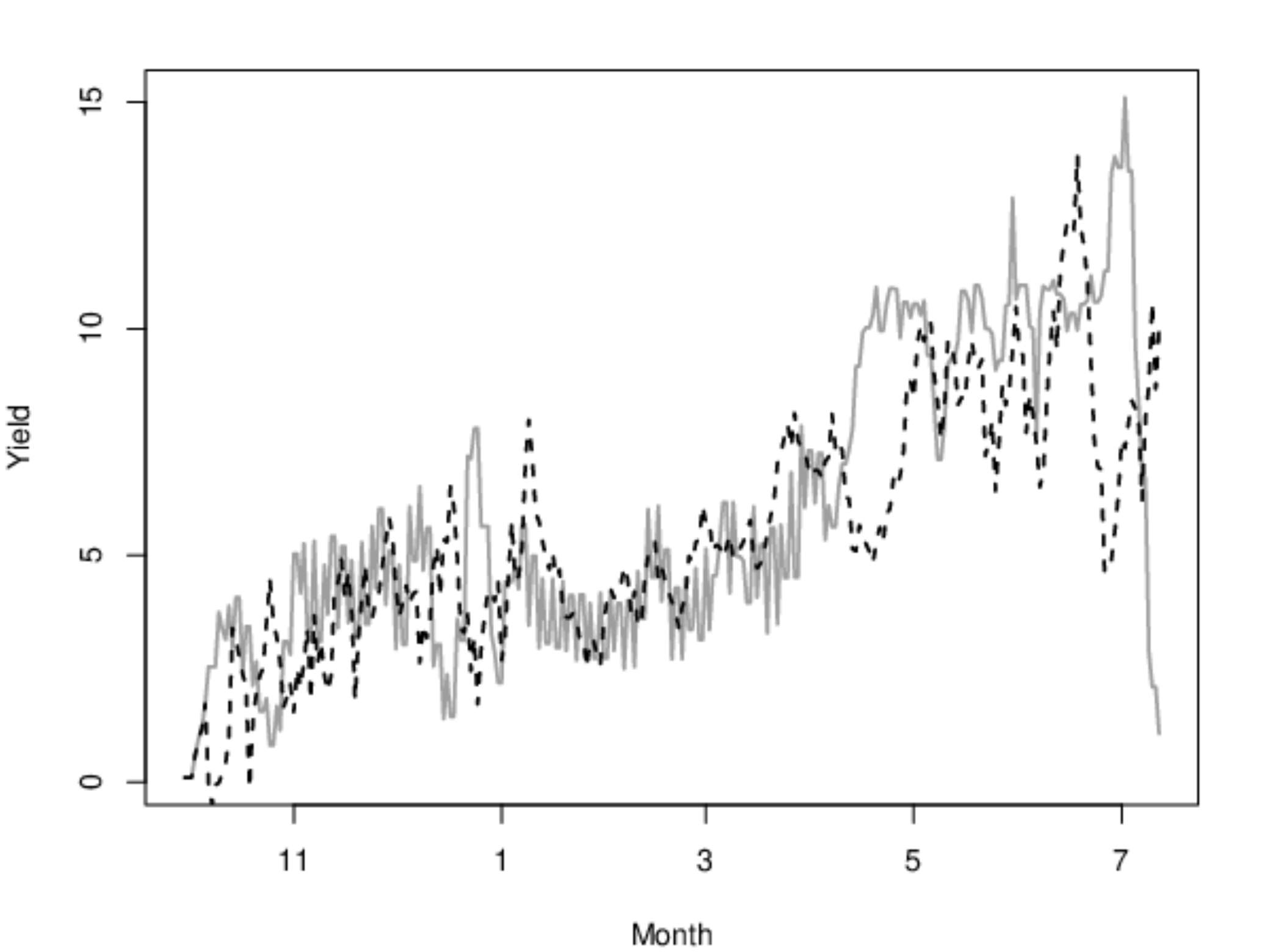}		
	\end{center}
	\caption{Prediction result for the crop yield data.  Gray curves show the data for the third period (gray) and black-dashed curve is the predicted curve.} 
	\label{fig:tomato_predict}
\end{figure}

We also performed the prediction of the crop yield.  
Here we predicted the weekly averaged yield for the $i_0+7$-th day using the data up to the $i_0$-th day, where $i_0$ is an index of the individual.    
The reason for predicting the crop yield for the $i_0+7$-th day instead of that for the $i_0+1$-th day is to prevent data leakage because the data for the response corresponding to $i_0$-th day consist of the average of the daily yield data from the $i_0$-th day to the $i_0+6$-th day.
First, the data corresponding to the first two periods $\{x_i(s), y_i; i=1,\ldots, i_0, i_0=550\}$ are used as training data.  
We applied the VCFAM to analyze this dataset, and then predicted the data for the $i_0+7$-th day $\{x_i(s), y_i; i=i_0+7\}$ as test data. 
We further repeated this analysis by incrementing $i_0$ to $n-7$ to predict the yield in the third period.  
Figure \ref{fig:tomato_predict} shows the crop yield data for the third period and their predictions.  
The prediction results roughly capture the trends in the data.  
However, in some points they give large prediction errors.  
The reason for it seems that there are variations that were not seen in the two periods used as the training set.  
We also tried to apply the VCFLM as a similar way, but failed to predict since the tuning parameters are not selected appropriately by the model selection criterion.  

\section{Concluding Remarks}
We have proposed a varying-coefficient functional additive model to capture the nonlinear structure of functional predictors and a scalar response varying with an exogenous variable.  
The proposed model is estimated by the penalized likelihood method, and then it is evaluated by a model selection criterion.  
Simulation studies show the effectiveness of our method in viewpoints of the prediction and the reconstruction of the effect of the predictor on the response.  
We also applied the proposed method to the analysis of crop yield data to investigate the effect of the environmental factors to the cultivation.  

In this work we considered the situation with a single functional predictor, while the future work include the extension to that with multiple functional predictors.
The crop yield is considered to be affected by not only the temperature but also several other environmental factors such as CO$_2$ concentration and solar radiations in the greenhouse.  
In this case, we also want to know which combination of environmental factors relates to the crop yield.  
To do so, the application of the sparse regularization \citep{HaTiWa2015} to the VCFAM is considered.  
It also remains as a future work to introducing sparsity inducing penalties to the VCFAM framework, using the idea of \cite{RaLaLi_etal2009} and \cite{MaKo2011}.  
\subsection*{Acknowledgments}
We would like to thank Higashibaba farm providing the data for cultivation of tomatoes.  
This work is supported by JSPS KAKENHI Grant Number 19K11858 and JST PRESTO Grant Number JPMJPR16O6.


\begin{thebibliography}{35}
	\newcommand{\enquote}[1]{``#1''}
	\expandafter\ifx\csname natexlab\endcsname\relax\def\natexlab#1{#1}\fi
	
	\bibitem[{Akaike(1974)}]{Ak1974}
	Akaike, H. (1974), \enquote{A new look at the statistical model
		identification,} \textit{IEEE Trans. Auto. Control}, 19, 716--723.
	
	\bibitem[{Cardot et~al.(1999)Cardot, Ferraty, and Sarda}]{CaFeSa1999}
	Cardot, H., Ferraty, F., and Sarda, P. (1999), \enquote{Functional linear
		model,} \textit{Statist. Probab. Lett.}, 45, 11--22.
	
	\bibitem[{Cardot and Sarda(2008)}]{CaSa2008}
	Cardot, H. and Sarda, P. (2008), \enquote{Varying-coefficient functional linear
		regression models,} \textit{Comm. Statist. Theory Methods}, 37, 3186--3203.
	
	\bibitem[{Davenport et~al.(2018)Davenport, Maity, and
		Baladandayuthapani}]{DaMaBa2018}
	Davenport, C.~A., Maity, A., and Baladandayuthapani, V. (2018),
	\enquote{{Functional interaction-based nonlinear models with application to
			multiplatform genomics data},} \textit{Stat. Med.}, 37, 2715--2733.
	
	\bibitem[{Fahrmeir et~al.(2013)Fahrmeir, Kneib, Lang, and
		Marx}]{FaKnLa_etal2013}
	Fahrmeir, L., Kneib, T., Lang, S., and Marx, B. (2013), \textit{Regression},
	Berlin Heidelberg: Springer.
	
	\bibitem[{Fan et~al.(2015)Fan, James, and Radchenko}]{FaJaRa2015}
	Fan, Y., James, G.~M., and Radchenko, P. (2015), \enquote{{Functional additive
			regression},} \textit{Ann. Statist.}, 43, 2296--2325.
	
	\bibitem[{Goldsmith et~al.(2010)Goldsmith, Feder, and Crainiceanu}]{GoFeCr2010}
	Goldsmith, J., Feder, J., and Crainiceanu, C. (2010), \enquote{Penalized
		functional regression,} \textit{J. Comput. Graph. Statist.}, 20, 830--851.
	
	\bibitem[{Green and Silverman(1994)}]{GrSi1994}
	Green, P. and Silverman, B. (1994), \textit{Nonparametric regression and
		generalized linear models: a roughness penalty approach}, London: Chapman \&
	Hall/CRC.
	
	\bibitem[{Han et~al.(2018)Han, M{\"{u}}ller, and Park}]{HaMuPa2018}
	Han, K., M{\"{u}}ller, H.~G., and Park, B.~U. (2018), \enquote{{Smooth
			backfitting for additive modeling with small errors-in-variables, with an
			application to additive functional regression for multiple predictor
			functions},} \textit{Bernoulli}, 24, 1233--1265.
	
	\bibitem[{Hastie and Tibshirani(1990)}]{HaTi1990}
	Hastie, T. and Tibshirani, R. (1990), \textit{Generalized additive models},
	London: Chapman \& Hall/CRC.
	
	\bibitem[{Hastie and Tibshirani(1993)}]{HaTi1993}
	--- (1993), \enquote{Varying-coefficient models,} \textit{J. Roy. Statist. Soc. Ser. B}, 55, 757--796.
	
	\bibitem[{Hastie et~al.(2015)Hastie, Tibshirani, and Wainwright}]{HaTiWa2015}
	Hastie, T., Tibshirani, R., and Wainwright, M. (2015), \textit{Statistical
		Learning with Sparsity: The Lasso and Generalization}, Boca Raton: Chapman \&
	Hall/CRC.
	
	\bibitem[{Hoover et~al.(1998)Hoover, Rice, Wu, and Yang}]{HoRiWu_etal1998}
	Hoover, D., Rice, J., Wu, C., and Yang, L. (1998), \enquote{Nonparametric
		smoothing estimates of time-varying coefficient models with longitudinal
		data,} \textit{Biometrika}, 85, 809--822.
	
	\bibitem[{Ivanescu et~al.(2015)Ivanescu, Staicu, Scheipl, and
		Greven}]{IvStSc_etal2015}
	Ivanescu, A.~E., Staicu, A.~M., Scheipl, F., and Greven, S. (2015),
	\enquote{Penalized function-on-function regression,} \textit{Comput. Statist.}, 30, 539--568.
	
	\bibitem[{Kokoszka and Reimherr(2017)}]{KoRe2017}
	Kokoszka, P. and Reimherr, M. (2017), \textit{Introduction to functional data
		analysis}, Boca Raton: CRC Press.
	
	\bibitem[{Konishi and Kitagawa(2008)}]{KoKi2008}
	Konishi, S. and Kitagawa, G. (2008), \textit{Information criteria and
		statistical modeling}, New York: Springer.
	
	\bibitem[{Li et~al.(2017)Li, Huang, and Zhu}]{LiHuZh2017}
	Li, J., Huang, C., and Zhu, H. (2017), \enquote{{A Functional
			Varying-Coefficient Single Index Model for Functional Response Data},}
	\textit{J. Am. Stat. Assoc.}, 1459, 0--0.
	
	\bibitem[{Matsui et~al.(2009)Matsui, Kawano, and Konishi}]{MaKaKo2009}
	Matsui, H., Kawano, S., and Konishi, S. (2009), \enquote{Regularized functional
		regression modeling for functional response and predictors,} \textit{J. Math-for-Industry}, 1, 17--25.
	
	\bibitem[{Matsui and Konishi(2011)}]{MaKo2011}
	Matsui, H. and Konishi, S. (2011), \enquote{Variable selection for functional
		regression models via the L1 regularization,} \textit{Comput. Statist. Data Anal.}, 55, 3304--3310.
	
	\bibitem[{Mclean et~al.(2014)Mclean, Hooker, Staicu, Scheipl, and
		Ruppert}]{McHoSt_etal2014}
	Mclean, M.~W., Hooker, G., Staicu, A.-m., Scheipl, F., and Ruppert, D.~R.
	(2014), \enquote{{Functional Generalized Additive Models Functional
			Generalized Additive Models},} \textit{J. Comput. Graph. Statist.}, 8600.
	
	\bibitem[{Morris(2015)}]{Mo2015}
	Morris, J.~S. (2015), \enquote{Functional regression,} \textit{Ann. Rev. Stat. Appl.}, 2, 321--359.
	
	\bibitem[{M{\"{u}}ller et~al.(2013)M{\"{u}}ller, Wu, and Yao}]{MuWuYa2013}
	M{\"{u}}ller, H.-G., Wu, Y., and Yao, F. (2013), \enquote{{Continuously
			additive models for nonlinear functional regression},} \textit{Biometrika},
	100, 607--622.
	
	\bibitem[{M{\"{u}}ller and Yao(2008)}]{MuYa2008}
	M{\"{u}}ller, H.-G. and Yao, F. (2008), \enquote{{Functional Additive Models},}
	\textit{J. Am. Stat. Assoc.}, 103, 1534--1544.
	
	\bibitem[{Peng et~al.(2016)Peng, Zhou, and Tang}]{PeZhTa2016}
	Peng, Q.~Y., Zhou, J.~J., and Tang, N.~S. (2016), \enquote{{Varying coefficient
			partially functional linear regression models},} \textit{Stat. Papers},
	57, 827--841.
	
	\bibitem[{Ramsay and Dalzell(1991)}]{RaDa1991}
	Ramsay, J. and Dalzell, C. (1991), \enquote{Some tools for functional data
		analysis,} \textit{J. Roy. Statist. Soc. Ser. B}, 53,
	539--572.
	
	\bibitem[{Ramsay and Silverman(2005)}]{RaSi2005}
	Ramsay, J. and Silverman, B. (2005), \textit{Functional data analysis (2nd
		ed.)}, New York: Springer.
	
	\bibitem[{Ravikumar et~al.(2009)Ravikumar, Lafferty, Liu, and
		Wasserman}]{RaLaLi_etal2009}
	Ravikumar, P., Lafferty, J., Liu, H., and Wasserman, L. (2009), \enquote{Sparse
		additive models,} \textit{J. Roy. Statist. Soc. Ser. B},
	71, 1009--1030.
	
	\bibitem[{Reiss et~al.(2017)Reiss, Goldsmith, Shang, and
		Ogden}]{ReGoSh_etal2017}
	Reiss, P.~T., Goldsmith, J., Shang, H.~L., and Ogden, R.~T. (2017),
	\enquote{{Methods for Scalar-on-Function Regression},} \textit{Int. Stat. Rev.}, 85, 228--249.
	
	\bibitem[{Scheipl et~al.(2016)Scheipl, Gertheiss, and Greven}]{ScGeGr2016}
	Scheipl, F., Gertheiss, J., and Greven, S. (2016), \enquote{{Generalized
			functional additive mixed models},} \textit{Electron. J. Stat.}, 10, 1455--1492.
	
	\bibitem[{Scheipl and Greven(2016)}]{ScGr2016}
	Scheipl, F. and Greven, S. (2016), \enquote{{Identifiability in penalized
			function-on-function regression models},} \textit{Electron. J. Stat.}, 10, 495--526.
	
	\bibitem[{Wong et~al.(2019)Wong, Li, and Zhu}]{WoLiZh2019}
	Wong, R.~K., Li, Y., and Zhu, Z. (2019), \enquote{{Partially Linear Functional
			Additive Models for Multivariate Functional Data},} \textit{J. Am. Stat. Assoc.},
	114, 406--418.
	
	\bibitem[{Wu et~al.(2010)Wu, Fan, and M{\"u}ller}]{WuFaMu2010}
	Wu, Y., Fan, J., and M{\"u}ller, H. (2010), \enquote{Varying-coefficient
		functional linear regression,} \textit{Bernoulli}, 16, 730--758.
	
	\bibitem[{Yao et~al.(2005{\natexlab{a}})Yao, M\"{u}ller, and
		Wang}]{YaMuWa2005a}
	Yao, F., M\"{u}ller, H., and Wang, J. (2005{\natexlab{a}}), \enquote{Functional
		data analysis for sparse longitudinal data,} \textit{J. Am. Stat. Assoc.}, 100,
	577--590.
	
	\bibitem[{Yao et~al.(2005{\natexlab{b}})Yao, M{\"u}ller, and
		Wang}]{YaMuWa2005b}
	Yao, F., M{\"u}ller, H., and Wang, J. (2005{\natexlab{b}}), \enquote{Functional
		linear regression analysis for longitudinal data,} \textit{Ann. Statist.}, 33, 2873--2903.
	
	\bibitem[{Zhu et~al.(2014)Zhu, Yao, and Zhang}]{ZhYaZh2014}
	Zhu, H., Yao, F., and Zhang, H.~H. (2014), \enquote{{Structured functional
			additive regression in reproducing kernel Hilbert spaces},} \textit{J. Roy. Statist. Soc. Ser. B}, 76,
	581--603.
	
\end{thebibliography}

\end{document}